\documentclass[twocolumn]{aastex631}

\DeclareUnicodeCharacter{202F}{\,}
\usepackage[T1]{fontenc}

\newcommand{\ltext}[1]{{\color{black}{\textbf\small{#1}}}}
\newcommand{\dtext}[1]{{\color{black}{\textbf\small{#1}}}}

\begin{document}

\title{Predicting Quasar Counts Detectable in the LSST Survey}

\correspondingauthor{Guodong Li, Roberto J. Assef}
\email{gdli@pku.edu.cn, roberto.assef@mail.udp.cl}

\author[0000-0003-4007-5771]{Guodong Li}
\affiliation{Kavli Institute for Astronomy and Astrophysics, Peking University, Beijing 100871, People's Republic of China}
\affiliation{National Astronomical Observatories, Chinese Academy of Sciences, 20A Datun Road, Beijing 100101, People's Republic of China}
\affiliation{Instituto de Estudios Astrof\'{i}sicos, Facultad de Ingenier\'{i}a y Ciencias, Universidad Diego Portales, Av. Ej\'{e}rcito Libertador 441, Santiago, Chile}

\author[0000-0002-9508-3667]{Roberto J. Assef}
\affiliation{Instituto de Estudios Astrof\'{i}sicos, Facultad de Ingenier\'{i}a y Ciencias, Universidad Diego Portales, Av. Ej\'{e}rcito Libertador 441, Santiago, Chile}

\author[0000-0002-0167-2453]{W.N. Brandt}
\affiliation{Department of Astronomy and Astrophysics, 525 Davey Lab, The Pennsylvania State University, University Park, PA 16802, USA}
\affiliation{Institute for Gravitation and the Cosmos, The Pennsylvania State University, University Park, PA 16802, USA}
\affiliation{Department of Physics, 104 Davey Laboratory, The Pennsylvania State University, University Park, PA 16802, USA}

\author[0000-0001-8433-550X]{Matthew J. Temple}
\affiliation{Centre for Extragalactic Astronomy, Department of Physics, Durham University, South Road, Durham DH1 3LE, UK}

\author[0000-0002-8686-8737]{Franz E. Bauer} 
\affiliation{Instituto de Alta Investigaci{\'{o}}n, Universidad de Tarapac{\'{a}}, Casilla 7D, Arica, Chile}

\author[0000-0002-1380-1785]{Marcin Marculewicz} 
\affiliation{Department of Physics and Astronomy, Wayne State University, 666 W. Hancock St, Detroit, MI, 48201, USA}

\author[0000-0002-5854-7426]{Swayamtrupta Panda}
\thanks{Gemini Science Fellow}
\affiliation{International Gemini Observatory/NSF NOIRLab, Casilla 603, La Serena, Chile}

\author[0000-0003-2196-3298]{Alessandro Peca} 
\affiliation{Eureka Scientific, 2452 Delmer Street, Suite 100, Oakland, CA 94602-3017, USA}
\affiliation{Department of Physics, Yale University, P.O. Box 208120, New Haven, CT 06520, USA}

\author[0000-0001-5231-2645]{Claudio Ricci} 
\affiliation{Instituto de Estudios Astrof\'{i}sicos, Facultad de Ingenier\'{i}a y Ciencias, Universidad Diego Portales, Av. Ej\'{e}rcito Libertador 441, Santiago, Chile}
\affiliation{Department of Astronomy, University of Geneva, ch. d'Ecogia 16, 1290, Versoix, Switzerland}

\author[0000-0002-1061-1804]{Gordon T. Richards}
\affiliation{Department of Physics, Drexel University, 32 S. 32nd Street, Philadelphia, PA, 19104 USA}

\author[0009-0003-0654-6805]{Sarath Satheesh-Sheeba} 
\affiliation{Instituto de Astrofísica, Facultad de Ciencias Exactas, Universidad Andrés Bello, Fernández Concha 700, 7591538 Las Condes, Santiago, Chile}

\author[0000-0002-9390-9672]{Chao-Wei Tsai}
\affiliation{National Astronomical Observatories, Chinese Academy of Sciences, 20A Datun Road, Beijing 100101, People's Republic of China}
\affiliation{Institute for Frontiers in Astronomy and Astrophysics, Beijing Normal University,  Beijing 102206, China}
\affiliation{University of Chinese Academy of Sciences, Beijing 100049, People's Republic of China}

\author[0000-0001-7808-3756]{Jingwen Wu}
\affiliation{University of Chinese Academy of Sciences, Beijing 100049, People's Republic of China}
\affiliation{National Astronomical Observatories, Chinese Academy of Sciences, 20A Datun Road, Beijing 100101, People's Republic of China}

\author[0000-0001-9163-0064]{Ilsang Yoon}
\affiliation{National Radio Astronomy Observatory, 520 Edgemont Rd, Charlottesville, VA 22903, USA}
\affiliation{Department of Astronomy, University of Virginia, P.O. Box 3818, Charlottesville, VA 22903, USA}



\begin{abstract}

The Legacy Survey of Space and Time (LSST), \ltext{being conducted} by the Vera C. Rubin Observatory, is a wide-field multi-band survey that will revolutionize our understanding of extragalactic sources through its unprecedented combination of area and depth. While the LSST survey strategy is still being finalized, the Rubin Observatory team has generated a series of survey simulations using the LSST Operations Simulator to explore the optimal survey strategy that best accommodates the majority of scientific goals. In this study, we utilize the latest simulated data and the Metrics Analysis Framework to predict the number of detectable quasars by LSST in each band and evaluate the impact of different survey strategies. We find that the number of quasars and lower luminosity AGNs detected in the baseline strategy (v4.3.1) in the redshift range $z$=0.3-6.7 will be highest in the \textit{i}-band (about 12 million) and lowest in the \textit{u}-band (about 6 million). Over 70\% of quasars are expected to be detected within the first year in all bands, as LSST will have already reached the break of the luminosity function at most redshifts.
\ltext{With a limiting magnitude of 25.7 (26.9) mag, we expect to detect 184 (199) million AGNs in the $z$-band ($r$-band) over the 10-year survey, with quasars constituting only 6\% of the total AGNs in each band. This arises because, considering that the luminosities of most low-luminosity AGNs are affected by contamination from their host galaxies, we set a magnitude threshold when predicting the number of quasars.} We find that variations in the \textit{u}-band strategy can impact the number of quasar detections. \ltext{Specifically, the difference between the baseline strategy and that with the largest total exposure in \textit{u} is 15\%.} In contrast, changes in rolling strategies, DDF strategies, weather conditions, and Target of Opportunity observations result in variations below 2\%. These results provide valuable insights for optimizing approaches to maximize the scientific output of quasar studies.

\end{abstract}


\section{Introduction} \label{sec:intro}

\begin{figure}[htbp]
\raggedright
\epsscale{1.2}
\plotone{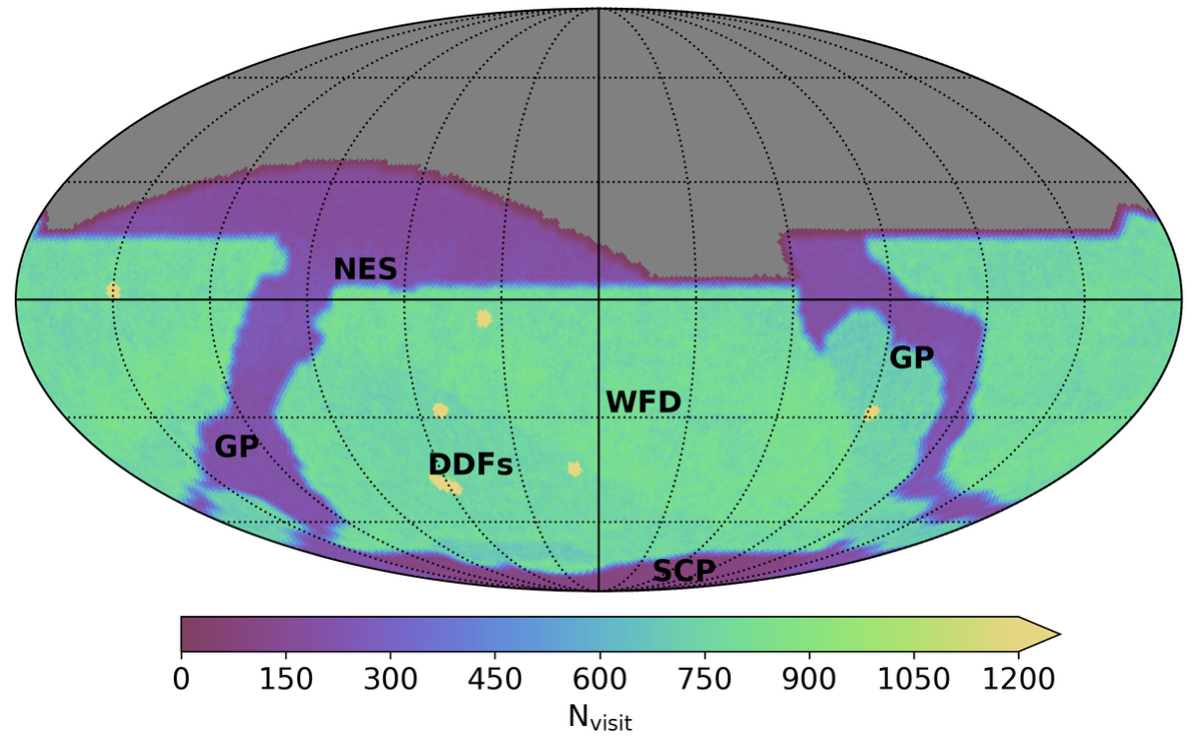}
\caption{LSST Ten-Year Survey footprint in equatorial coordinates, showing the number of visits ($\rm N_{visit}$) conducted per sky location (generated from Baseline v4.3.1). The light purple areas correspond to the mini-surveys, including the Galactic plane and polar regions as described in Section \ref{sec:intro}. The cyan area represents the WFD region ($\sim$20,000 deg$^2$). See SRD for more detail.
\label{fig:fig1}}
\end{figure}

\ltext{A comprehensive census of the population of active galactic nuclei (AGN) is essential for understanding the growth of supermassive black holes (SMBHs) and the formation and evolution of galaxies \citep[][]{2004MNRAS.351..169M,2012NewAR..56...93A,2014ARA&A..52..415M,2015ARA&A..53...51S}, as the growth of the central SMBH is closely linked to the evolution of its host galaxy, particularly through AGN feedback on the star formation activity \citep{2012ARA&A..50..455F,2024Galax..12...17H}. As AGN display a wide range of observational properties, studies of accretion activity ranging from low luminosity AGNs to the brightest quasars are paramount to achieve these goals.

Observational astronomy is entering a new era, characterized by the advent of a new generation of telescopes designed for wide and deep sky surveys, with the Legacy Survey of Space and Time (LSST) at the forefront of this revolution \citep[e.g.,][]{2009arXiv0912.0201L,2019ApJ...873..111I,2022ApJS..258....1B}.} The LSST to be conducted by the Vera C. Rubin Observatory, is a ten-year survey project aimed at repeatedly imaging the visible sky above Cerro Pachón in northern Chile in 6 broad-band filters (\textit{ugrizy}) in the 320-1050 nm wavelength range \citep{2009arXiv0912.0201L}. The survey is obtained by the 8.36-meter (6.5-meter effective aperture) Simonyi Survey Telescope, whose camera \citep[LSSTCam;][]{2018SPIE10705E..0DR} features a circular field of view (FOV) of 9.6 square degrees. LSST is scheduled to begin scientific observations in \ltext{early 2026}, covering the southern sky with a median single-visit depth of $r\sim$ 24.3. The combination of wide sky coverage, significant depth, and repeated observations enables the project to pursue several major scientific goals, such as studying the Solar System, transient and variable objects, the Milky Way, and the physics of dark matter and dark energy \citep{lsst2004,2019ApJ...873..111I}.

The LSST will be conducted as a collection of multiple surveys operating in parallel (see Figure \ref{fig:fig1}). Its primary component is the Wide-Fast-Deep (WFD) survey, a wide-area survey covering $\sim$ 20,000 square degrees of the sky with a uniform observing strategy. This survey will consist of a median total of $\sim$825$\times$30s exposures divided across six filters over the 10-year period. Approximately 90\% of the observing time will be dedicated to the WFD survey. The co-added observation depths will reach between 25.1 and 27.7 mag depending on the band \citep{2017arXiv170804058L, 2019ApJ...873..111I}. More detailed requirements for the WFD are outlined in the LSST Science Requirements Document (SRD)\footnote{The LSST Science Requirements Document is available at \url{https://ls.st/srd.}}. The remaining time will be distributed among a variety of mini-surveys and micro-surveys ($\sim$ 5\% of the time), as well as $\sim$ 5\% dedicated to Deep Drilling Fields (DDFs), which involve more frequent observations of a small number of fixed locations centered on fields with significant additional data: the Cosmic Evolution Survey field \citep[COSMOS; e.g.,][]{2007ApJS..172....1S,2016ApJ...819...62C}, the Extended Chandra Deep-Field South field \citep[ECDFS; e.g.,][]{2002ApJS..139..369G,2005ApJS..161...21L,2022hxga.book...78B},  the Euclid Deep-Field South \citep[EDFS; e.g.,][]{2011arXiv1110.3193L,2019sptz.prop14235S}, the European Large Area ISO Survey-S1 field \citep[ELAIS-S1; e.g.,][]{2000MNRAS.316..749O,2000MNRAS.316..768S,2000MNRAS.319.1169E,2001ApJ...554...18A,2022hxga.book...78B}, and the XMM-Large Scale Structure field \citep[XMM-LSS; e.g.,][]{2004JCAP...09..011P,2018MNRAS.478.2132C}.

These surveys will produce a database containing observations of $\sim$20 billion galaxies and a similar number of stars \citep{2019ApJ...873..111I}, revolutionizing our understanding of the dynamic universe, particularly in the fields of extragalactic astronomy and AGN \citep[e.g.,][]{2017arXiv170804058L, 2018arXiv180901669T, 2018ApJS..236....9N, 2018arXiv180901669T, 2018arXiv181106542B, 2021ApJS..253...31L, Savic}. \ltext{For example, the use of LSST to study AGN will significantly improve constraints on AGN population demographics and luminosity functions, thereby enhancing our understanding of the accretion history of SMBHs across cosmic time. It will also provide new constraints on the physics and structure of AGN and their accretion disks, as well as the mechanisms by which gas is supplied to sustain their growth. Additionally, given the high-redshift reach of LSST, it will enable the study of SMBH formation and co-evolution with their host galaxies and dark matter halos over much of cosmic history.} 

\begin{figure*}[htbp]
\raggedright
\epsscale{1.1}
\plotone{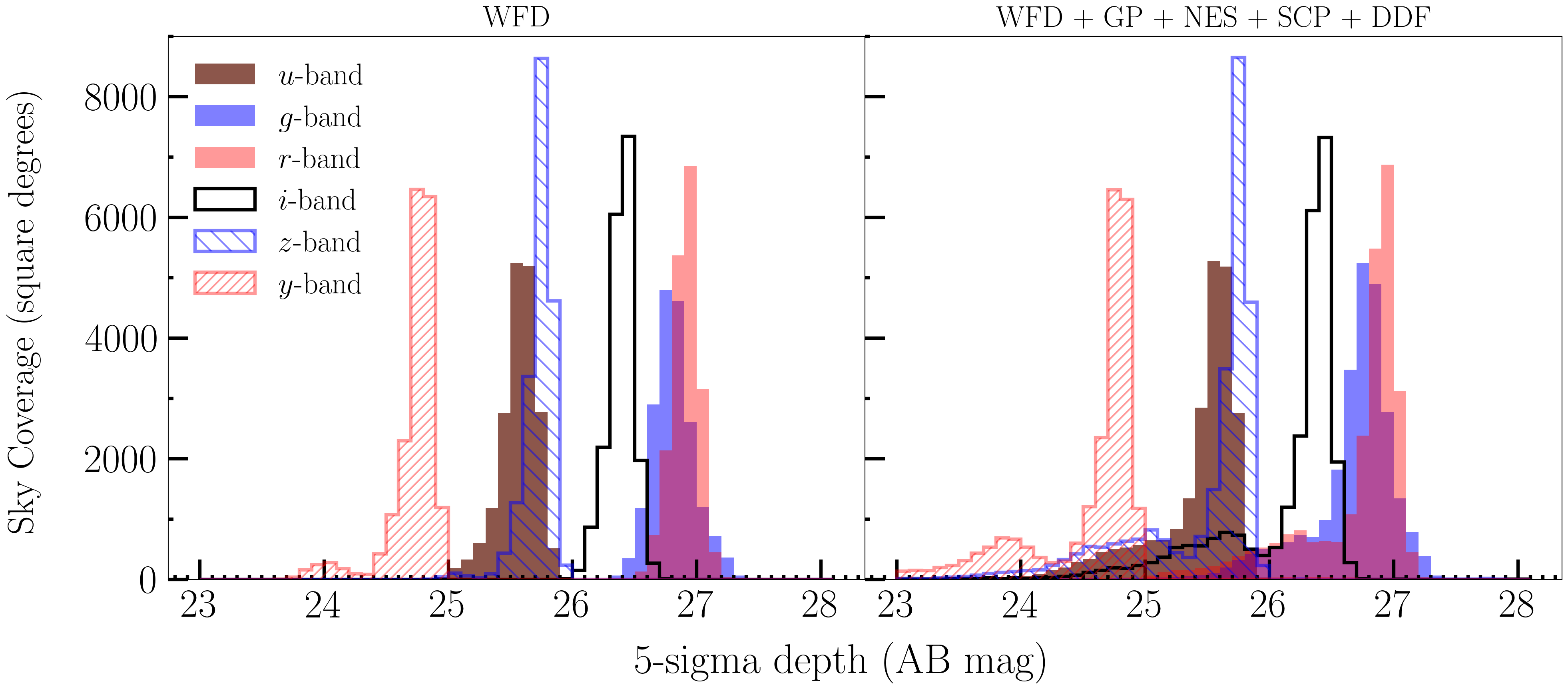}
\caption{Histograms of the 5$\sigma$ depth (AB mag) for the \textit{u}, \textit{g}, \textit{r}, \textit{i}, \textit{z}, and \textit{y} bands based on the LSST ten-year survey simulation data. Left: Expected 5$\sigma$ depth for Wide Fast Deep (WFD) survey region. Right: Expected 5$\sigma$ depth for regions observed by all LSST sub-surveys, namely the WFD, Galactic Plane (GP), South Celestial Pole (SCP), North Ecliptic Spur (NES), and Deep Drilling Fields (DDFs).
\label{fig:fig2}}
\end{figure*}

The observation strategy and cadence for LSST have not yet been fully decided \citep[e.g.,][]{2017arXiv170804058L,2022ApJS..258....1B,2023ApJS..266...22S}. The Survey Cadence Optimization Committee (SCOC)\footnote{\url{https://pstn-056.lsst.io/}} will continue to adjust the cadence based on the results from the first year of the LSST survey, in order to meet the requirements of time-domain, cosmology, extragalactic science, and other LSST scientific goals. To explore various options and their impact on the main scientific goals of the survey, the Rubin Observatory LSST Scheduler Team has developed the LSST Operations Simulator \citep[OpSim\footnote{\url{https://www.lsst.org/scientists/simulations/opsim}};][]{2014SPIE.9150E..14C,2014SPIE.9150E..15D,2017arXiv170804058L,jones2020} using the scheduler \citep[\texttt{rubin\_sim/OpSim\footnote{\url{https://github.com/lsst/rubin_sim}}};][]{2019AJ....157..151N,2022zndo...7087823Y}, which combines science program requirements, telescope design mechanics, and environmental condition modeling to provide a framework for operational simulations. This framework simulates the telescope movements, observing conditions and image characteristics, and simulations are periodically released by the Vera C. Rubin Observatory.\footnote{\url{https://survey-strategy.lsst.io}} 

In this work, we take into account the latest LSST baseline v4.3.1 observing strategy to estimate the number of actively accreting super-massive black holes identifiable in the LSST survey. \ltext{To align with the LSST Science Book \citep{lsstsciencecollaboration2009lsstsciencebookversion}, we focus on estimating the number of high-luminosity objects with $M_{\rm i}$$<$-20 mag. We note that we have also made simpler estimates for lower luminosity objects based on significantly different assumptions. For clarity, in our study, the former are referred to as quasars, as their luminosities are consistent with those of quasars, at least in the local universe \citep{2024A&A...683A.112S, 2025arXiv250504259S}, while the latter are referred to as AGN, although we recognize the arbitrariness of this definition.} We quantify the impact of variations of this strategy on quasar detection based on the LSST Metrics Analysis Framework \citep[MAF;][]{2014SPIE.9149E..0BJ}. In Section 2, we outline the MAF adopted for this study and describe the quasar and AGN number calculations. The results are discussed in Section 3. We summarize our work in Section 4. In this paper, we adopt a flat $\Lambda$CDM cosmology with $H_{\rm 0} \rm{=70~km~s^{-1}~{Mpc}^{-1}~and~\Omega_m=0.3}$.

\begin{deluxetable*}{cccccccccc}[htbp]
\tablecaption{Number of Quasars detected in different regions of the LSST survey \label{tab:table1}}
\tablewidth{0pt}
\tablehead{
\colhead{Filter} & \colhead{WFD} & \colhead{GP} & \colhead{SCP} & \colhead{NES} &  \colhead{DDFs} & \colhead{N$_{\rm QSO}$ (total)}  & \colhead{Area (deg$^2$)} & \colhead{5$\sigma$ depth (median)} & \colhead{N$_{\rm QSO/deg^{2}}$ (median)}
}
\startdata
u & 86.65\% & 8.15\% & 3.18\% & 1.94\% &  0.08\%& 6461417  & 23754  & 25.5 mag & 253\\
g & 75.37\% & 11.57\% & 3.12\% & 9.90\% & 0.04\% & 10496226  & 26530  & 26.7 mag & 349\\
r & 75.24\% & 11.87\% & 2.97\% & 9.88\% & 0.04\% & 11949081  & 26565  & 26.9 mag & 398\\
i & 75.40\% & 11.81\% & 2.90\% & 9.84\% &  0.05\% & 12260573  & 26573  & 26.4 mag & 408\\
z & 76.02\% & 11.70\% & 2.70\% & 9.50\% & 0.08\% & 11883070  & 26553  & 25.7 mag & 399\\
y & 85.02\% & 10.25\% & 2.79\% & 1.86\% & 0.08\% & 9387109  & 23706 & 24.7 mag & 357\\
\enddata
\tablecomments{The WFD (Wide Fast Deep), GP (Galactic Plane), SCP (South Celestial Pole), NES (North Ecliptic Spur), and DDFs (Deep Drilling Fields) represent different regions within the LSST survey, as shown in Figure \ref{fig:fig1}. The percentages indicate the fraction of QSOs that can be detected in each region across different bands, relative to the total number of QSOs ($\rm N_{QSO}$). We also show the coverage area, the median 5$\sigma$ depth, and the median $\rm N_{QSO}$ detected per square degree across different bands in the LSST survey.}
\end{deluxetable*}
\vspace{-1.0cm}

\section{Method} \label{sec:method}

\subsection{Metrics Analysis Framework} \label{subsec:maf}

The Python-based Metrics Analysis Framework \citep[MAF\footnote{\url{https://www.lsst.org/scientists/simulations/maf}};][]{2014SPIE.9149E..0BJ} is a tool used to read and evaluate simulated observations generated by OpSim. Specifically, OpSim produces a simulation product containing various parameters describing each visit (time, pointing, filter, field of view, limiting magnitude), while MAF (\texttt{rubin\_sim/Maf}) uses slicers and metrics to derive plots and statistics that help evaluate specific science goals in the simulated survey. Slicers are used to define subsets (``slices") of the simulated data according to specific needs, and metrics calculate values for each slice that relate to the specific science goals\footnote{To see an example of how to use the slice and metric: \url{https://github.com/lsst/rubin_sim_notebooks/tree/main/maf/tutorial}}. Additionally, MAF accepts SQL commands for constraints, allowing us to select specific components from the simulated data such as specific dates, to analyze.

To estimate the number of quasars, we use the \texttt{QSONumberCountsMetric}, described in the next section, while to estimate the number of AGN we use a variation of this metric as described in Section \ref{subsec:AGN}. They both rely on the \texttt{ExgalM5} MAF metric, which provides the effective 5$\sigma$ depth of the survey at a given position in the sky by applying foreground reddening estimates from the dust maps of \citet{1998ApJ...500..525S} and \citet{2011ApJ...737..103S} to the nominal 5$\sigma$  depth estimated with the \texttt{Coaddm5Metric} metric. We divide the sky in HEALPix \citep{2005ApJ...622..759G} cells with NSIDE=64 using the MAF provided slicer \texttt{HealpixSlicer}. Each HEALPix cell has an approximate size of 55$^{\prime}$, well below the LSST Camera FoV (9.6 $\rm deg^{2}$).  Figure \ref{fig:fig2} shows the expected distribution of effective 5$\sigma$ depths of the HEALPix cells for the WFD, DDFs, and the combined mini and micro surveys in baseline v4.3.1.

\subsection{Quasar Counts Metric} \label{subsec:metric}

\texttt{QSONumberCountsMetric} is a metric that estimates the number of quasars per square degree in the different bands of the LSST survey for a given slicer pixel by integrating the observed quasar luminosity function (QLF). Specifically, we use observed QLFs derived from the work of \citet{2020MNRAS.495.3252S}, who used a double power-law parameterization to study the bolometric QLF over the redshift range 0$-$7:
\begin{equation}\label{eqn:QLF}
\Phi(L, z) = \frac{\Phi_*(z)}{(L/L_*(z))^{\gamma_1(z)} + (L/L_*(z))^{\gamma_2(z)}},
\end{equation}
where $\Phi_*(z)$ is the quasar co-moving space number density for a given bolometric luminosity $L$ at a redshift $z$, $L_*(z)$ is the break luminosity, and $\gamma_1(z)$ and $\gamma_2(z)$ are the faint-end and bright-end slopes, respectively. The \texttt{QSONumberCountsMetric} is based on the Model ``A" QLF provided by \citet{2020MNRAS.495.3252S}. Following \citet{2020MNRAS.495.3252S}, we can convert the bolometric luminosity function into the intrinsic (i.e., reddening corrected) luminosity function at specific wavelength ($\lambda = \lambda_{\rm eff}/(1+z)$), assuming the quasar SED template of \citet{2006ApJS..166..470R} and the bolometric corrections derived by \citet{2020MNRAS.495.3252S}. We then apply the reddening distribution of \citet{2014ApJ...786..104U} in the same manner as done by \citet{2020MNRAS.495.3252S} to derive observed QLF, $\Phi_{\rm Obs}$. The reddening distribution is also consistent with those from larger-area surveys, such as Stripe 82X \citep{2016ApJ...817..172L,2024HEAD...2110206L}. Notably, as discussed by \citet{2020MNRAS.495.3252S}, the $N_{\rm H}$ distribution model from \citet{2014ApJ...786..104U} may input an uncertainty of $\sim$ 0.2 dex in the binned estimations of the bolometric QLFs. Additionally, we include absorption by the intergalactic medium (IGM) using the model from \citet{2014MNRAS.442.1805I}. We only consider objects with an intrinsic absolute \textit{i}-band magnitude $M_i<-20~\rm mag$ to avoid lower luminosity AGN whose observed fluxes may be dominated by the host galaxy, emulating the cut applied in the LSST Science Book \citep{lsstsciencecollaboration2009lsstsciencebookversion} for an initial estimation (see Section \ref{subsec:AGN} for estimates of the lower luminosity objects we refer to as AGN). 

\ltext{For $z < $0.3, the volume is too small to contain a sufficient number of quasars for the luminosity functions to be robust. At $z > $6.7, Lyman-$\alpha$ forest absorption shifts into the $y$-band, resulting in single-band detections that provide inadequate information for study with LSST. In this work, we adopt the redshift range of $z = 0.3$ to 6.7, consistent with the range used in a number of related scientific projects outlined in the LSST Science Book \citep{lsstsciencecollaboration2009lsstsciencebookversion}.}
We also consider observed magnitudes between the nominal LSST saturation limit (14.3, 15.8, 15.8, 15.6, 15.1, 14.1 for the \textit{u}, \textit{g}, \textit{r}, \textit{i}, \textit{z}, \textit{y} bands, respectively) down to the 10-year stack 5$\sigma$ limiting magnitude for the corresponding HEALPix cell. The saturation magnitudes correspond to the median values output by the \texttt{QSONumberCountsMetric} and \texttt{ExgalM5} for the baseline v4.3.1 observing strategy. The magnitude limits in a given position are provided by the \texttt{ExgalM5} MAF metric, restricted to regions with foreground extinction \ltext{$E(B-V)<1.0$} to avoid highly dust-obscured areas near the Galactic Plane and Galactic Center. The survey regions used in this work include the WFD survey area, DDFs, and several mini-surveys \ltext{(Galactic Plane-GP, North Ecliptic Spur-NES, and South Celestial Pole-SCP)}, covering a total of $\sim$ 27,000 square degrees.

\ltext{To further explore the impact of the adopted QLF on the predicted quasar counts, we extended our analysis to include the QLFs presented by \citet{2007ApJ...654..731H} and \citet{2011ApJ...728...56A}. Relative to the predictions based on \citet{2020MNRAS.495.3252S}, the counts derived 
from \citet{2007ApJ...654..731H} are approximately 1.5 times higher in all bands, while those from \citet{2011ApJ...728...56A} mid-infrared QLF are about two times lower across all bands. A detailed description of the assumptions and resulting predictions is provided in the Appendix.
Given that the QLF model in \citet{2020MNRAS.495.3252S} is based on a more extensive compilation of observational data—spanning infrared, B-band, ultraviolet, and both soft and hard X-ray observations over the past few decades—we adopt the QLF from \citet{2020MNRAS.495.3252S} for the subsequent analyses in this study.}


\begin{figure}[htbp]
\raggedleft
\epsscale{1.1}
\plotone{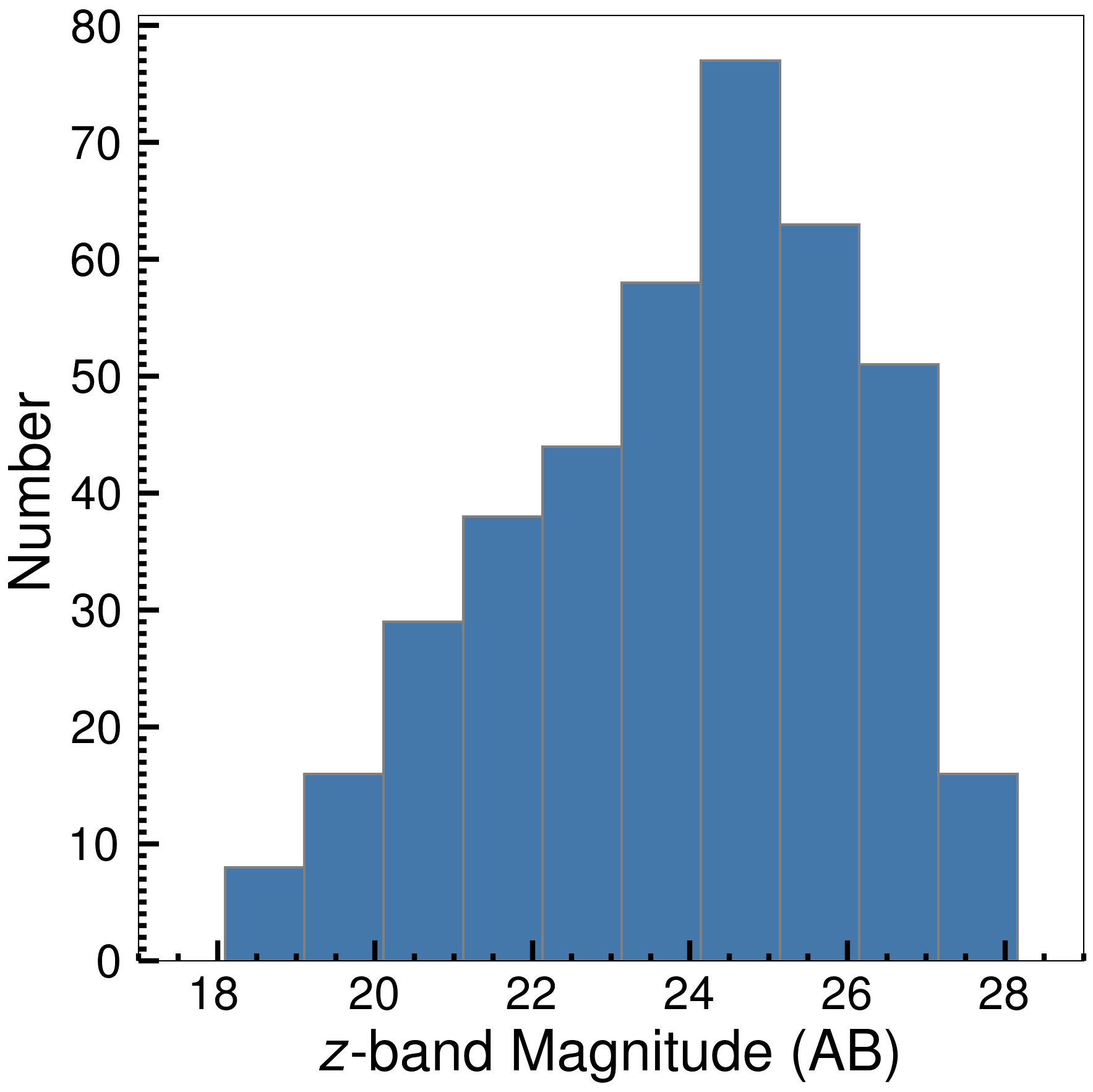}
\caption{The $z$-band magnitude distribution of AGN in the GOODS region (160 arcmin$^2$) from the catalog of \citet{2017ApJS..228....2L}. 
\label{fig:AGN}}
\end{figure}

\subsection{AGN Counts Estimation} \label{subsec:AGN}

Estimating the number of detectable lower luminosity AGN is fundamentally different than for quasars, as the luminosity function across cosmic time is not well constrained. So, instead of using a model, we rely on measured number counts of AGNs in deep, pencil-beam field observations. Specifically, we use the number counts per square degree as a function of magnitude, in combination with the survey depth in each HEALPix cell. For each cell, the \texttt{ExgalM5} MAF metric is used to determine the 5$\sigma$ depth, as described in Section \ref{subsec:maf}. The total number of detectable AGN is obtained by summing the contributions from all cells, where each cell (with an area of $\sim$0.8 $\rm deg^2$) contributes based on the integration of the AGN number counts up to its 5$\sigma$ depth. 

In this work, we adopt the AGN number count distribution from \citet{2017ApJS..228....2L}, who identified and compiled an AGN catalog using X-ray data from the Chandra Deep Field-South (CDF-S) combined with multi-wavelength analysis. This catalog covers an area of 484.2 arcmin$^2$, with the highest AGN density reaching $\approx$23,900 $\rm deg^{-2}$, which is consistent with the lower end of the log $N$ - log $S$ relation observed in Stripe 82XL \citep{2024ApJ...974..156P}. And its completeness levels are 74.7\%, 90.5\%, and 62.8\% in the full, soft, and hard X-ray bands, respectively \citep{2017ApJS..228....2L}. To obtain the optical magnitude distribution of AGNs, we select the Great Observatories Origins Deep Survey (GOODS) South region (160 arcmin$^2$; see Figure 1 in \citealp{2017ApJS..228....2L}) within the CDF-S, a contiguous area characterized by deep optical detection \citep[$m_{\rm 5\sigma}$=28.2 mag;][]{2004ApJ...600L..93G}. The magnitude distribution of these AGNs is shown in Figure \ref{fig:AGN}. \ltext{We note that the predicted number of AGN, along with a comparison to the number of AGN in larger fields, is discussed in detail in Section \ref{subsec:AGN1}.}

\begin{figure*}[htbp!]
\raggedleft
\epsscale{1.1}
\plotone{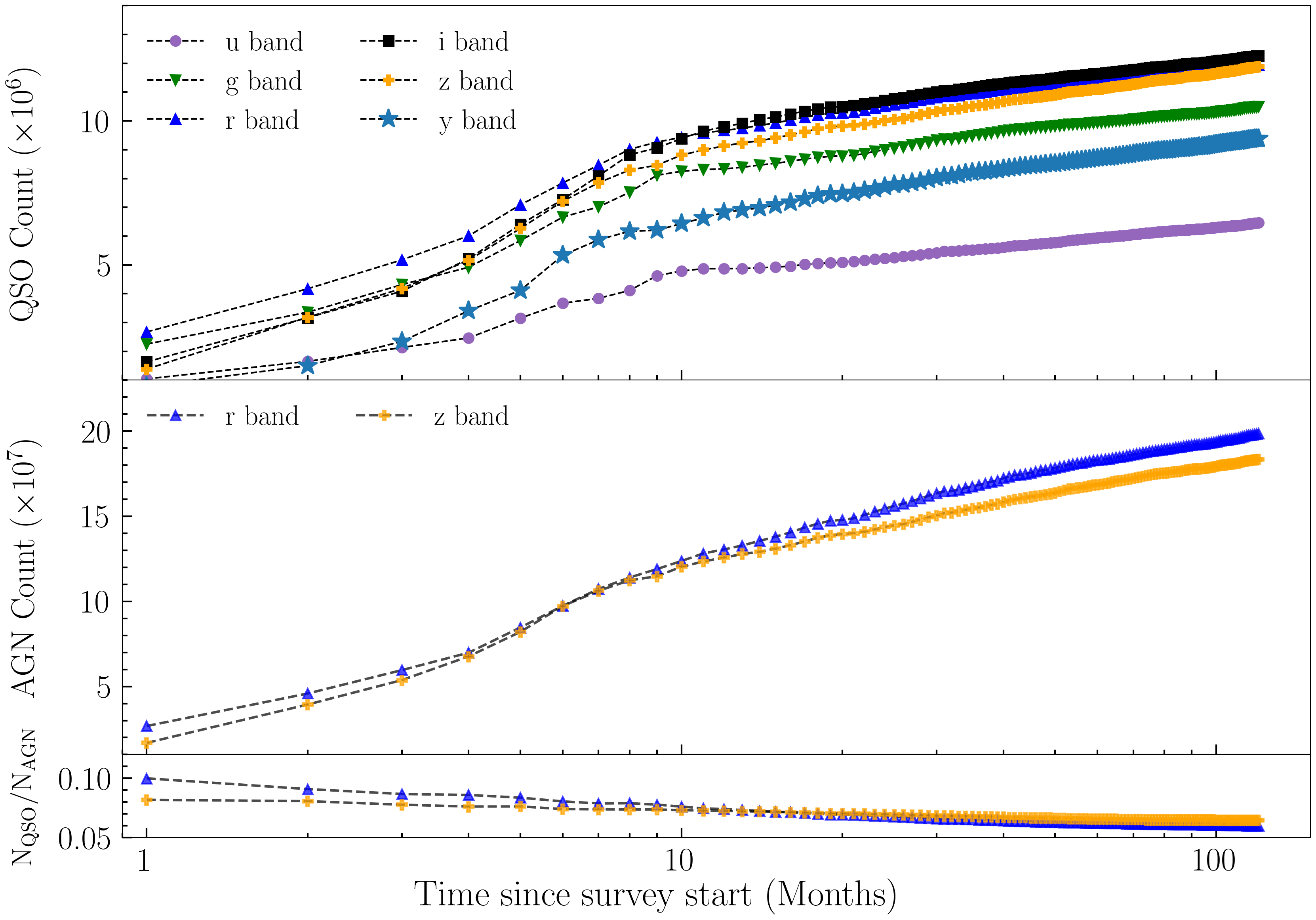}
\caption{\textbf{Upper}: The monthly variation of quasar counts in optical bands (\textit{ugrizy}) over the ten-year period in the LSST survey. \textbf{Middle}: The monthly variation of AGN counts in $r$ and $z$ band over the ten-year period from 2025 to 2035 in the LSST survey. \textbf{Lower}: Monthly variation in the number of quasar detections relative to AGN detections.
\label{fig:fig3}}
\end{figure*}

\section{Result} \label{sec:QSO_N}

\subsection{Baseline Cadence} \label{subsec:baseline}

As mentioned earlier, the LSST Survey Scheduling Team regularly releases simulations created using OpSim that respond to recommendations from the SCOC. Each release provides a simulation with a baseline strategy that is designed to meet the LSST survey requirements, as described in \citet{2018arXiv180901669T}. Each release also includes simulations with minor adjustments to the baseline strategy to explore changes that could optimize different science goals \citep{2014SPIE.9150E..14C, 2014SPIE.9150E..15D, 2017arXiv170804058L, jones2020}. As of the writing of this article, the most up-to-date version of the simulated strategies is version 4.3.1. However, the most extensive testing for strategy adjustments, particularly for \textit{u}-band and rolling cadences, was conducted on version 3.4\footnote{https://community.lsst.org/t/release-of-v3-4-simulations/8548}\ltext{, with subsequent baseline strategies being continuously updated and adjusted based on it.} In comparison, the v4.3.1 baseline focuses more on practical operations, incorporating considerations for downtime, start dates, and maintenance schedules to improve the realism of the simulations (see \href{https://survey-strategy.lsst.io/index.html}{Vera C. Rubin Observatory Survey Strategy} for more detail). Therefore, in our work, we use the baseline 4.3.1 strategy along with other strategies from version 3.4.

\subsubsection{Quasar Count} \label{subsec:QSO}

Using the metric discussed in Section \ref{subsec:metric}, we estimate that the number of quasars detectable (defined as those with a significance greater than 5$\sigma$) in the baseline observing strategy of the {\texttt v4.3.1} release ranges from $\sim$6 million in the $u$-band to $\sim$12 million in the $i$-band, as shown in Table \ref{tab:table1}. Note that not all bands will cover the same total area, with $g$-, $r$-, $i$-, and $z$-band covering $\sim$10\% more area than $u$ and $y$. The \textit{i}-band detects the most quasars with an estimated number of 12.3 million. The \textit{g}- and \textit{r}-band have similar survey areas and comparable depth to the \textit{i}-band, but the number of detected quasars is expected to be somewhat lower, 10.5 and 12.0 million respectively, due to dust obscuration and IGM absorption. The \textit{u}-band detects the fewest quasars, as it is the most affected by dust obscuration and IGM absorption, and has a smaller sky coverage. Interestingly, despite having a smaller sky coverage and a relatively shallow depth compared to the \textit{u}-band, the \textit{y}-band detects 45\% more quasars, reaching 9.4 million, as it is less affected by dust obscuration and is not subject to IGM absorption within the redshift range considered.

\begin{figure*}[htbp]
\raggedleft
\epsscale{1.1}
\plotone{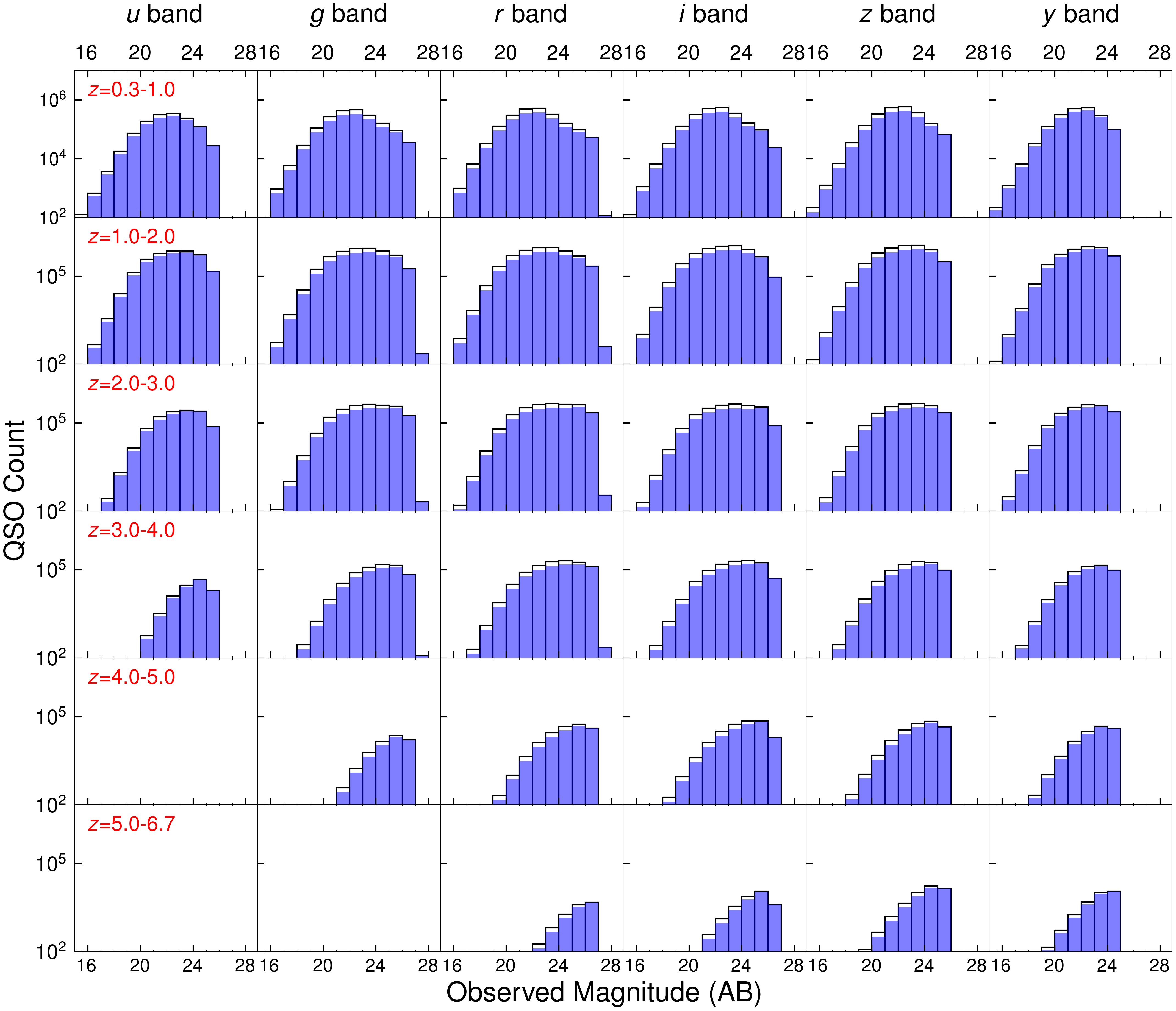}
\caption{The distribution of quasar counts as a function of observed magnitude (AB) in different redshift ranges and filters. Each column represents a specific filter ($u$, $g$, $r$, $i$, $z$, and $y$ from left to right), and each row corresponds to a redshift range. \ltext{The black hollow bins denote the number of quasars detected in each magnitude bin across all regions in the LSST survey (WFD+GP+SCP+NES+DDFs), with the blue bins corresponding to the WFD.}
\label{fig:fig4}}
\end{figure*}

By slicing the simulated data from 2025 to 2035 into monthly segments, we explore the variation in the number of quasars detected across different bands over the survey period.  By the end of the first survey year, we expect to detect over 70\% of quasars across all optical bands (see Figure \ref{fig:fig3}). Following Equation 14 from \citet{2020MNRAS.495.3252S} to calculate the break luminosity of the QLF at a given redshift and convert it to the observed magnitude, we confirm that the LSST survey depth can reach this break luminosity at most redshifts ($z$$\sim$0.3-6.0) as early as the end of the first year. \dtext{We note that although the $r$-band detects the most quasars in the first year, the $i$-band ultimately detects a higher number of quasars due to the narrowing of the survey depth difference (which shrinks from $\sim$ 0.8 to $\sim$ 0.5 mag) over time, combined with the lower efficiency of the r-band in detecting objects at $z>$ 3 due to IGM absorption (see Figure \ref{fig:fig4}).}

Figure \ref{fig:fig4} shows the total quasar counts over a ten-year period as a function of observed magnitude in different redshift ranges. Quasars at  $z>$ 3 are challenging to detect in the \textit{u}-band due to the effects of the IGM absorption. Similarly, the \textit{g}-band is also affected by IGM absorption, particularly at $z > 4$. In contrast, more $z > 4$ quasars are detectable in other bands (\textit{rizy}), as those bands are less affected by IGM absorption. Among these, the \textit{r}- and \textit{i}-bands, with their larger survey areas and greater depths (see Table \ref{tab:table1}), are expected to detect more high-$z$ quasars compared to other bands (\textit{ugzy}). Additionally, quasar counts in all bands are concentrated around $z\sim$1–2, which aligns with the peak of the quasar luminosity density in this redshift range \citep[e.g.,][]{1984ApJ...283..486M,1994MNRAS.271..639B,1997MNRAS.285..547J,2000A&A...353...25M,2005A&A...441..417H,2009A&A...493...55E,2011ApJ...728...56A,2015ApJ...804..104M,2016A&A...587A.142F,2020MNRAS.495.3252S}.

\begin{figure*}[htbp]
\raggedleft
\epsscale{1.1}
\plotone{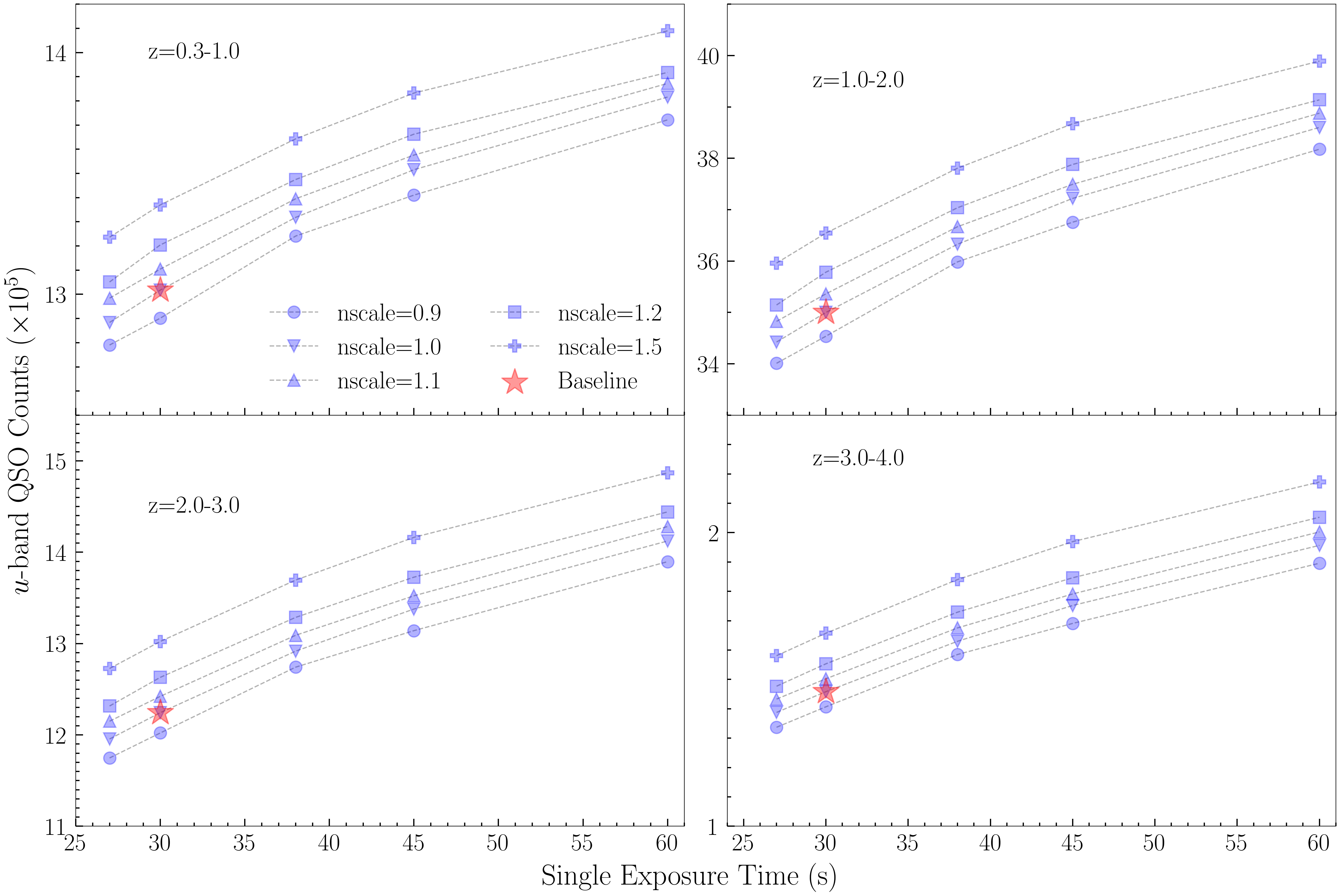}
\caption{Quasar counts for the \textit{u}-band in the LSST survey as a function of single exposure time. The panels show the quasar counts for different redshift ranges: $z$=0.3-1.0 (top-left), $z$=1.0-2.0 (top-left), $z$=2.0-3.0 (bottom-left), and $z$=3.0-4.0 (bottom-right). The red star indicate the baseline Quasars counts for a 30-second exposure time with \texttt{nscale}=1.0. Different lines represent different \texttt{nscale} values, which are multiples of $\rm N_{visit}$ for the baseline strategy .
\label{fig:fig5}}
\end{figure*}

\subsubsection{AGN Count} \label{subsec:AGN1}

Using the approach described in Section \ref{subsec:AGN}, we estimate that up to 184 million AGNs can be detected in the $z$-band across the entire LSST area. The number of identified quasars estimated using the QLF method (see Section \ref{subsec:metric}) is only 6\% of the total AGN population (see Figure \ref{fig:fig3}). More than 70\% of these AGNs are expected to be detected within the first year. \ltext{Using the \textit{WISE} AGN catalog in \citep{2018ApJS..234...23A},} we find that $\sim$ 2.6 million objects from this catalog lie within the LSST/WFD survey area, accounting for $\sim$ 2\% of the AGNs expected to be identified in this region. \ltext{Using optical variability to select AGN, \citet{2021A&A...645A.103D} found that the number density of AGN is $\sim$ 344 $\rm deg^{-2}$, predicting about 9.29 million AGN in the LSST area, which accounts for about 5\% of our predicted AGN count.}

\ltext{There is a significant variation in the number density of AGN detected across different X-ray energy bands. At the soft band flux ($\sim 4.2\times 10^{-18} \rm \ erg \ cm^{-2} \ s^{-1}$), \citet{2017ApJS..228....2L} found the AGN density to be $\sim$ 23,900 $\rm deg^{-2}$, which is consistent with the lower end of the log $N$ - log $S$ relation observed in Stripe 82XL \citep{2024ApJ...974..156P}. Assuming the highest AGN density, we expect $\sim$ 635 million AGN to be detected in the LSST $z$-band survey. In our work, we predict the number of AGN in the LSST area to be $\sim$184 million based on the optical magnitude distribution of AGN detected in X-rays (see Section \ref{subsec:AGN}; \citealp{2017ApJS..228....2L}). Given that 45\% of the AGNs in the CDF-S region from \citet{2017ApJS..228....2L} lack spectroscopic redshifts, we estimate the relative cosmic variance for this region to be $\sim$ 6\%, based on the measurements from \citet{2010MNRAS.407.2131D} in the GEMS survey \citep{2008ApJS..174..136C}. This result also applies to our estimate of the AGN counts from the LSST survey affected by cosmic variance, as we are only scaling the total numbers. We note that, due to the incompleteness of high-redshift galaxy samples, this estimate should be treated as a reference (see \citealp{2010MNRAS.407.2131D} for more detail).}

A similar analysis was conducted for the $r$ band, where we cross-matched with the CANDELS/GOODS catalog \citep{2011ApJS..197...35G,2011ApJS..197...36K} to obtain \textit{HST}/F606W photometry, yielding results consistent with those in the $z$-band. Specifically, we estimate that up to 199 million AGNs could be detected in the $r$-band, of which 6\% are quasars. The undetected AGN are likely low-luminosity systems such as dwarf AGNs \citep[e.g.,][]{2025ApJ...982...10P}, potentially obscured by contamination from their host galaxies, and are therefore not accounted for in our $\rm N_{quasar}$ predictions (see Section \ref{subsec:metric}). We note that the $z$-band AGN magnitude distribution we used is from the GOODS-S catalog \citep{2004ApJ...600L..93G}, with a 5$\sigma$ limiting magnitude of 28.2 mag, which includes 84\% of the AGNs detected in the GOODS-S region through X-ray approach by \citet{2017ApJS..228....2L}. This result suggests that the AGN count estimated in our work may be underestimated.

\subsection{u-band Observation Cadence} \label{subsec:u-band}

The \textit{u}-band filter allocation is decoupled from the regular cadence of other filters \citep{2018arXiv181200515L}. The depth of the \textit{u}-band survey is crucial for detecting key spectral features like the Lyman break, particularly for galaxies in the redshift range of 2-3, which corresponds to the cosmic noon, and for improving photometric redshift estimates \citep[e.g.,][]{2019arXiv190410439C, crenshaw2025quantifyingimpactlsstuband}. Given the potential importance of \textit{u}-band for such science goals, a number of different alternative strategies are considered in the release of the scheduler simulations. In the OpSim simulations 3.4, there are 25 strategies for the \textit{u}-band, which vary the single frame exposure times (specifically 27s, 38s, 45s, and 60s, compared to the 30s of the baseline strategy) and the number of visits (see Figure \ref{fig:fig5}). The simulations specifically consider \texttt{nscale} values of 0.9, 1.1, 1.2, and 1.5, \ltext{corresponding to strategies with fewer or more visits per region relative to the baseline strategy.}

Given the higher optical depth of dust in the $u$-band, increasing exposure time or the number of visits (``\texttt{nscale}") \ltext{is not expected to} significantly affect quasar detection. Moreover, under the baseline strategy, the \textit{u}-band depth can quickly reach the break luminosity within the first year for most redshifts, while the flatter part of the luminosity function shows a slower increase in number counts with depth. \ltext{With adjustments to the exposure time and number of visits (\texttt{nscale}), the total number of quasars detected in the \textit{u}-band ranges from 5,988,579 to 7,102,423 (Figure \ref{fig:fig5}). Compared to the \textit{i}-band, which detects the highest number of quasars (baseline strategy), the \textit{u}-band detects only 58\% of the quasars, even with the longest exposure and maximum visits (60s and 1.5 \texttt{nscale}). When compared to the \textit{u}-band baseline strategy, increasing the exposure time and the number of visits for low-redshift quasars ($z = 0.3 - 1.0$) can improve detection by up to 8\% (60s and 1.5 \texttt{nscale}). For high-redshift quasars ($z = 1.0 - 4.0$), increasing the exposure time by a factor of 1.5 leads to a $\sim$7\% increase in detection, while increasing the number of visits by the same factor results in about a 5\% improvement.}


\begin{figure*}[htbp]
\raggedleft
\epsscale{1.1}
\plotone{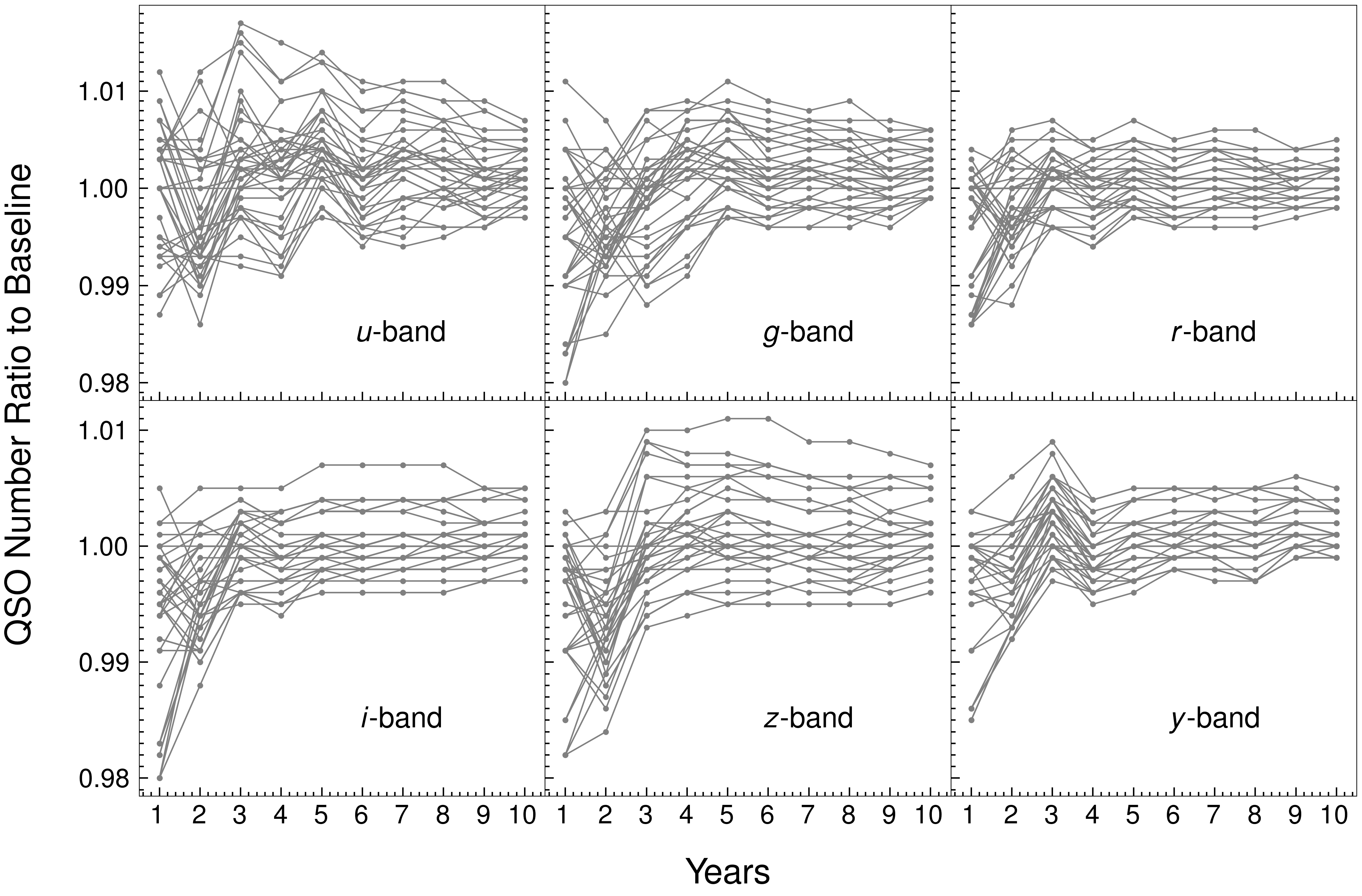}
\caption{Annual trends in quasar number ratios for different rolling strategies to the baseline strategy in the LSST survey. Each panel represents one of the six bands (\textit{ugrizy}), showing the predicted quasar numbers over ten years for \ltext{40 rolling cadences}. The x-axis indicates the year of observation, while the y-axis represents the ratio of quasar numbers for different rolling strategies to the baseline strategy (see Figure \ref{fig:fig3}). Each line represents a different rolling strategy, as detailed in Section \ref{subsec:rolling}.
\label{fig:fig6}}
\end{figure*}

\begin{figure*}[htbp]
\raggedleft
\epsscale{1.1}
\plotone{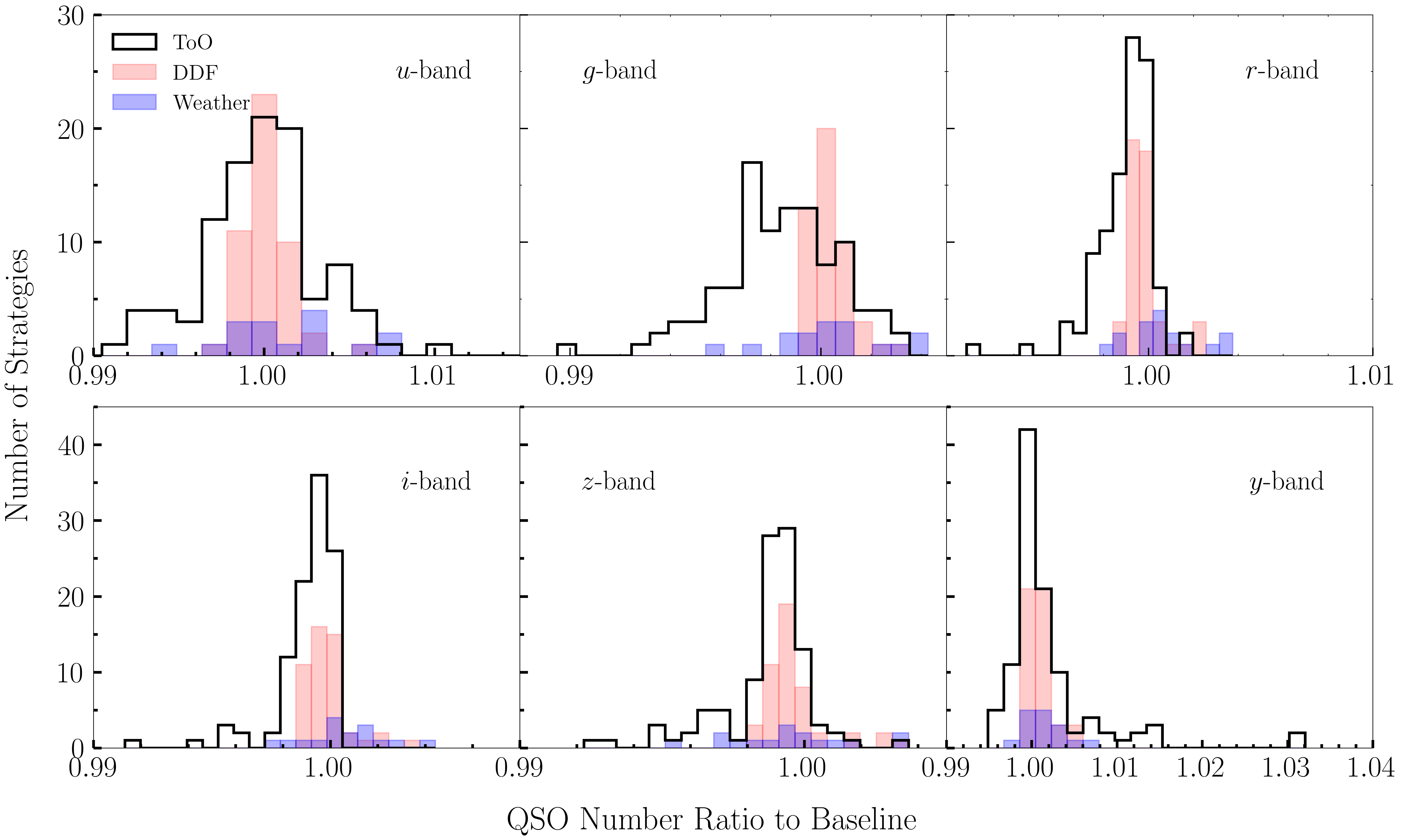}
\caption{Distribution of quasar number ratios for different strategies to the baseline strategy over the 10-year LSST survey. Each subplot represents a different band. The black line is the ToO (Target of Opportunity) strategies, the blue represents ``weather" related strategies, and the red represents DDF strategies.
\label{fig:fig7}}
\end{figure*}

\subsection{Rolling Cadence \label{subsec:rolling}}

Rolling cadence refers to an observing strategy where different regions of the sky are prioritized for visits at different times, allowing for increased observation frequency in specific areas while maintaining balanced overall coverage. In the baseline strategy (v4.3), a uniform rolling cadence with three cycles is adopted, with planned pauses in years 4 and 7, ultimately achieving uniform visit coverage. The simulations also consider alternate strategies that utilize variations in the details of the rolling cadence to enhance the coverage and temporal sampling of specific sky regions. Specifically, these simulations include scenarios with no rolling strategy at different starting times, rolling with different cycles, rolling with pauses at different starting times, uniform rolling starting at different times, and concentrated uniform rolling during the early stages at different starting points. In total, 40 different strategy simulations were generated.

To explore their impact on quasar detection, we applied the same analysis used for the Baseline strategy to these 40 rolling strategies. Figure \ref{fig:fig6} shows the number of quasars detected in different optical bands under these 40 strategies, along with their annual variations. Compared to the Baseline results, although the adjustments to the rolling strategies have an impact on quasar detection numbers during the first two years of the survey—resulting in up to a 2\% variation—the overall change in the total number of quasars over the entire ten-year period remains within 1\%. Therefore, adjustments to the rolling strategy do not have a significant impact on quasar detection. The impact, however, may be very significant for AGN variability studies as more intensive cadences improve the mapping of the structure function and probe crucial scales for photometric reverberation mapping measurements (see \citealp{kovacevic2021lsstagnsccadence}, \citealp{2022ApJS..262...49K} and the LSST AGN SC Cadence Note by Yu \& Richards 2021\footnote{\url{https://docushare.lsst.org/docushare/dsweb/Get/Document-37656/AGN_SF_Metric.pdf}} for further discussion).




\subsection{Other Cadence Variations} \label{subsec:others}

In addition to the adjustments to the \textit{u}-band and rolling strategies, the LSST Scheduler Team has conducted extensive tests on various observational strategies, focusing primarily on the DDF regions, weather conditions, and Target of Opportunity (ToO) observations. The DDFs are a set of carefully selected regions of the sky that are observed with higher cadence and depth compared to the main survey. These fields are crucial for calibrating photometric redshifts, improving weak-lensing measurements, and enabling transients and time-domain studies of quasars or AGNs \citep[e.g.,][]{2022ApJS..262...49K,2023PASP..135j5002H,2024ApJS..275...21G}. The simulations for different weather scenarios are based on potential variations in cloud cover and atmospheric conditions over the ten-year survey period. The ToO strategies are designed to optimize the ability to rapidly follow up on transient events, enhancing LSST responsiveness to such occurrences. These simulations include considerations of duration, cadence, and different band combinations (e.g., \textit{griz}, \textit{ugrizy}), which can affect the sky coverage and detection depth in the respective bands, thereby impacting the overall quasar detection of the survey.

The quasar number distribution over the ten-year period for these strategies (see Figure \ref{fig:fig7}) show that the variations in most bands (\textit{ugriz}) are within 2\% of the baseline strategy. Adjustments to the strategies related to the DDF regions, given their small area (total area $<$120 deg$^2$), have a minimal impact on the overall quasar counts, with changes of less than 1\%. We note that in the \textit{y}-band, the ToO strategies (``too\_elab\_r256\_d2\_ugrizy\_v3.4\_10yrs" and ``too\_elab\_r256\_d3\_ugrizy\_v3.4\_10yrs") cover an additional 2000 deg$^{2}$ compared to the baseline strategies, resulting in a 5\% increase in quasar detections. \ltext{Overall, strategies that explore different parameter spaces for the DDF regions, weather conditions, and ToO observations have less than a 2\% impact on the quasar count predicted using the QLF.}


\section{CONCLUSIONS} \label{sec:dis}

In this study, we estimated the number of AGNs and quasars that will be detected in the LSST survey. Using the \texttt{QSONumberCountsMetric}, we also evaluate the potential impact of various LSST survey strategies on quasar detection, based on the QLF from \citet{2020MNRAS.495.3252S}. We focus on the latest baseline strategy and the most extensively adjusted strategies. Specifically, we use version 4.3.1 for the baseline strategy, while other strategies are based on version 3.4, which includes the most adjustment tests and has generated the corresponding strategies (\textit{u}-band cadence, rolling strategies, DDFs, weather conditions and ToO observations).

\begin{itemize}

\item [1)] 
For the baseline v4.3.1 strategy, we predict that 6 to 12 million quasars can be detected within the six LSST bands. The \textit{i}-band is expected to detect the highest number of quasars due to its wide coverage and deep survey depth, while the \textit{u}-band, affected by dust obscuration and limited sky coverage, is expected to detect the fewest quasars. Based on the AGN distribution from \citet{2017ApJS..228....2L}, we estimate that up to 184 and 199 million AGNs can be detected in the $z$- and $r$-band respectively, with quasars accounting for approximately 6\% of the AGN population in the $r$- and $z$-band, likely due to contamination from their host galaxies.

\item [2)] 
By segmenting the simulated data of baseline strategy into monthly parts, we found that over 70\% of quasars would be detected \ltext{(and thus available to be identified)} in the first year, indicating that LSST will reach the break luminosity of luminosity function at most redshifts ($z$$\sim$0.3-6.0) within the first year. Most of the quasars detected in each band are fainter than 20 mag and are concentrated around $z\sim$1-2. \ltext{Due to the larger survey depth in the \textit{r}-, and \textit{i}-band, we anticipate that these bands will capture more quasars in the $z\sim2-6.7$ range compared to others.}

\item [3)]
The \textit{u}-band survey strategy is particularly important for improving the accuracy of photometric redshifts and for detecting high-redshift Lyman-break galaxies, which enable probes of the quasar environments. For our work, increasing exposure time and the number of visits slightly improves quasar detection, with a maximum increase of 8\% for low-redshift quasars ($z =$ 0.3-1) and 7\% for high-redshift quasars. These improvements are limited by dust obscuration and host galaxy contamination. Adjustments to the rolling strategy show minimal variations in quasar detection, with an impact of $<$ 1\% over the entire ten-year period compared to the baseline strategy. This indicates that rolling strategies do not significantly affect quasar detection.

\item [4)]
In the OpSim simulations v3.4, a total of 215 different survey strategies were produced. Excluding the \textit{u}-band and rolling-related strategies, \ltext{remaining strategies (i.e., those that explore different parameter spaces for the DDF regions, weather conditions, and for ToO observations), show an impact of less than 2\% on quasar detection in most bands, as estimated using the \citet{2020MNRAS.495.3252S} QLF, with no significant effects observed.}

\end{itemize}

Although the LSST survey strategy has not yet been finalized, our analysis of all considered survey strategies indicates that they do not significantly impact quasar detection compared to the baseline strategy. We quantified the variations in quasar detection within different redshift intervals and annually, providing a detailed understanding of these changes. Future LSST data releases will enable us to quickly evaluate and validate our results.

This work was supported by the National Natural Science Foundation of China (No. 11991052) and the China Manned Space Program (No. CMS-CSST-2025-A09). G.D.L. thanks the support from the Ministry of Science and Technology of the People's Republic of China (No. 2022YFA1602902) and the International Partnership Program of the Chinese Academy of Sciences (No.114A11KYSB20210010).
R.J.A. was supported by FONDECYT grant number 1231718 and by the ANID BASAL project FB210003. SP is supported by the international Gemini Observatory, a program of NSF NOIRLab, which is managed by the Association of Universities for Research in Astronomy (AURA) under a cooperative agreement with the U.S. National Science Foundation, on behalf of the Gemini partnership of Argentina, Brazil, Canada, Chile, the Republic of Korea, and the United States of America. MJT acknowledges funding from a FONDECYT Postdoctoral fellowship (3220516) and from STFC (ST/X001075/1). WNB acknowledges support from USA NSF grant AST-2407089. CR acknowledges support from SNSF Consolidator grant F01$-$13252, Fondecyt Regular grant 1230345, ANID BASAL project FB210003 and the China-Chile joint research fund.

This publication makes use of simulation databases from the Vera C. Rubin Observatory Teams, which are available at \url{https://s3df.slac.stanford.edu/data/rubin/sim-data/}. We acknowledge the use of code for this work, which is available at \url{https://github.com/guodongli2024/QSO-Counts-in-the-LSST-Survey}.


\vspace{5mm}
\software{lsst/rubin\_sim \citep{2022zndo...7087823Y}
          }

\ltext{

\section*{\textbf{Appendix}}
\subsection*{\rm Predicted Quasar Counts in the LSST Survey Using Alternative QLFs}

\begin{deluxetable*}{cccccccccc}[tbp]

\tablecaption{Number of Quasars detected in the LSST survey \label{tab:table2}}
\tablewidth{0pt}
\tablenum{2}
\tablehead{\colhead{Filter} & \colhead{WFD} & \colhead{GP} & \colhead{SCP} & \colhead{NES} &  \colhead{DDFs} & \colhead{N$_{\rm QSO}$ (total)}  & \colhead{Area (deg$^2$)} & \colhead{5$\sigma$ depth (median)} & \colhead{N$_{\rm QSO/deg^{2}}$ (median)}
}
\startdata
\multicolumn{10}{c}{ Quasar Luminosity Function from \citet{2007ApJ...654..731H}} \\ 
u & 88.22\% & 7.05\% &   2.90\% & 1.74\% & 0.09\% & 9095715 & 23754 & 25.5 mag & 108\\
g & 76.41\% & 10.99\% & 3.04\% & 9.52\% & 0.04\% & 16785193 & 26530 & 26.7 mag & 156\\
r & 76.10\% & 11.40\% & 2.86\% & 9.61\% & 0.03\% & 19404019 & 26565 & 26.9 mag & 183\\
i & 76.57\% & 11.29\% & 2.73\% & 9.37\% & 0.04\% & 19433442 & 26573 & 26.4 mag & 191\\
z & 77.47\% & 11.05\% & 2.52\% & 8.89\% & 0.07\% & 18102404 & 26553 & 25.7 mag & 191\\
y & 86.29\% & 9.41\% & 2.47\% & 1.72\% & 0.11\% & 13433683 & 23706 & 24.7 mag & 177\\
\hline
\multicolumn{10}{c}{ Quasar Luminosity Function from \citet{2011ApJ...728...56A}} \\ 
u & 86.13\% & 8.61\% & 3.26\% & 1.95\% & 0.05\% & 3929152 & 23754 & 25.5 mag & 153\\
g & 74.34\% & 12.20\% & 3.20\% & 10.24\% & 0.02\% & 6068461 & 26530 & 26.7 mag & 200\\
r & 74.33\% & 12.37\% & 3.06\% & 10.21\% & 0.03\% & 7255840 & 26565 & 26.9 mag & 239\\
i & 74.13\% & 12.56\% & 3.03\% & 10.25\% & 0.03\% & 6807585 & 26573 & 26.4 mag & 224\\
z & 74.74\% & 12.36\% & 2.89\% & 9.97\% & 0.04\% & 6596879 & 26553 & 25.7 mag & 218\\
y & 84.00\% & 11.07\% & 2.93\% & 1.94\% & 0.06\% & 5409719 & 23706 & 24.7 mag & 203\\
\enddata
\end{deluxetable*}

In our analysis, we primarily adopted the QLF from \citet{2020MNRAS.495.3252S}. To assess the impact of different QLFs on our results, we extended the estimation to incorporate the QLFs of \citet{2007ApJ...654..731H} and \citet{2011ApJ...728...56A}, combined with the baseline strategy, to predict the detected quasar counts in the LSST survey.


The QLF from \citet{2007ApJ...654..731H} was constructed over the redshift range $z=0$-6 using a comprehensive compilation of quasar data from optical, X-ray, and infrared surveys. This QLF differs from that of \citet{2020MNRAS.495.3252S} in several key ways. While \citet{2020MNRAS.495.3252S} follows the methodology of \citet{2007ApJ...654..731H}, their QLF features steeper bright- and faint-end slopes at high redshift ($z \gtrsim$ 2-3), which evolve with redshift. Another important distinction is that \citet{2007ApJ...654..731H} used a constant comoving number density normalization, whereas \citet{2020MNRAS.495.3252S} allowed this normalization to vary with redshift. Following the same assumptions adopted for \citet{2020MNRAS.495.3252S}, we used the SED template from \citet{2006ApJS..166..470R}. However, considering that the QLF from \citet{2007ApJ...654..731H} is based on the reddening distribution from \citet{2003ApJ...598..886U}, we adopt the same reddening distribution here for consistency with \citet{2007ApJ...654..731H}. Using the \texttt{QSONumberCountsMetric} described in Section~\ref{subsec:metric}, we estimate that in the redshift range $z=0.3$-6.0, LSST would detect quasar counts ranging from $\sim$9,095,715 in the $u$ band to $\sim$19,433,442 in the $i$ band (Table~\ref{tab:table2}), which are higher by a factor of $\sim$ 1.4-1.6 compared to the results derived using the QLF from \citet{2020MNRAS.495.3252S} over the same redshift range, and are consistent with the estimates in the LSST Science Book \citep{lsstsciencecollaboration2009lsstsciencebookversion}. We note that if we were to use the reddening distribution from \citet{2014ApJ...786..104U}, the predicted quasar counts would decrease by a factor of about 4.

Based on an AGN sample selected in the mid-infrared and X-ray, \citet{2011ApJ...728...56A} constructed the $J$-band QLF over the redshift range $0 < z < 5.85$, using a double power-law model and correcting the host-galaxy contamination with low-resolution empirical templates from \citet{2010ApJ...713..970A}. Here, we adopted the luminosity-and-density evolution (LDE) model given in Table 3 of \citet{2011ApJ...728...56A}, converted the $J$-band QLF to other bands with the SED template of \citet{2006ApJS..166..470R}.
Using the \texttt{QSONumberCountsMetric}, we estimate that in the redshift range $0.3 < z < 5.85$, LSST would detect $\sim$ 3,929,152 quasars in the $u$ band and 7,255,840  in the $r$ band (Table~\ref{tab:table2}), which are lower by a factor of $\sim$1.6–1.8 compared with predictions based on the QLF of \citet{2020MNRAS.495.3252S} over the same range. Compared with \citet{2020MNRAS.495.3252S}, the comoving number density normalization also evolves with redshift, while its bright- and faint-end slopes are re assumed to be redshift-independent.

Overall, both \citet{2007ApJ...654..731H} and \citet{2011ApJ...728...56A} yield quasar counts that are systematically offset by approximately a factor of $\sim$2 from those of \citet{2020MNRAS.495.3252S}, with \citet{2007ApJ...654..731H} yielding higher counts and \citet{2011ApJ...728...56A} yielding lower counts. These differences are primarily due to variations in QLF parameters. Constraining these parameters more tightly will require deeper and wider surveys in the future, such as LSST.


}

\bibliography{references}{}
\bibliographystyle{aasjournal}





\end{document}